\documentclass[epj,twocolumn]{webofc}
\usepackage[varg]{txfonts}   
\woctitle{18th International Symposium on Very High Energy Cosmic Ray Interactions}
\begin{document}
\title{Modelling hadronic interactions in HEP MC generators}
%
%

\author{Peter Skands\inst{1,2}\fnsep\thanks{\email{peter.skands@monash.edu}} 
}

\institute{Theory Unit, CERN, CH-1211 Geneva 23, Switzerland
\and
          School of Physics and Astronomy, Monash University,
          Melbourne, Australia 
          }

\abstract{%
HEP event generators aim to describe high-energy collisions in full
exclusive detail. They combine perturbative matrix elements and parton
showers with dynamical models of less well-understood phenomena such
as hadronization, diffraction, and the so-called underlying event. We
briefly summarise some of the main concepts relevant to the modelling
of soft/inclusive hadron interactions in MC generators, in particular
PYTHIA, with emphasis on questions recently highlighted by LHC data.
}
\maketitle
\section{Introduction}
\label{intro}
This summary was written in the context of the 18th
International Symposium on Very High Energy Cosmic Ray Interactions
(ISVHECRI).  It is based on two recent proceedings-style mini-reviews of 
soft-inclusive event-generator models\footnote{The first was 
a Snowmass / FCC-hh study~\cite{Skands:2013asa,Campbell:2013qaa,Barletta:2014vea} 
focusing on extrapolations of soft-physics models to 100 TeV CM energies. 
The other was a contribution on the PYTHIA generator(s) to a CERN
Yellow Report prepared by the LHC forward physics study group; in
progress.}, updated and extended in a
hopefully reasonably coherent and useful form. In section \ref{sec-SoftModels},
we give an overview of soft physics models, focusing on multi-parton
interactions (MPI). In section \ref{sec-ColRec}, we discuss the physics
of colour
reconnections (CR) along with some alternative proposed interpretations of observed properties of
particle spectra in hadron collisions. Finally in section \ref{sec-PyTunes},
we give a brief overview of the most recent tuning efforts in the
context of the PYTHIA 8 event generator, in particular the  so-called Monash 2013 tune. 

\section{Soft Physics Models}
\label{sec-SoftModels}
Soft physics models can essentially be divided into two broad
categories. The first starts from perturbative QCD (partons, matrix
elements, jets) and uses a factorized
perturbative expansion for the hardest parton-parton interaction, 
combined with parton showers and detailed models of hadronization and (soft and hard) 
multiparton interactions (MPI). This is the 
approach taken by general-purpose event generators,
like HERWIG~\cite{Corcella:2000bw,Bahr:2008pv},
PYTHIA~\cite{Sjostrand:2006za,Sjostrand:2014zea}, and
SHERPA~\cite{Gleisberg:2008ta}. Since they agree with perturbative QCD
(pQCD) at high $p_\perp$, they are used extensively by the
collider-physics community, 
see \cite{Buckley:2011ms,Skands:2012ts} for reviews. 
The price is a typically low predictivity for very soft
physics, though the modelling of diffractive and other soft-inclusive
phenomena is generally improving, and is an active area of research in all the generators. 
Collisions involving nuclei with $A\ge 2$ are generally not
addressed at all by these generators, though extensions
exist~\cite{Armesto:2009fj,Gyulassy:1994ew}.  

At the other end of the spectrum are tools starting from Regge
theory (optical theorem, cut and uncut pomerons), like
QGSJET~\cite{Ostapchenko:2004ss}. A priori, there are no jets
whatsoever in this formalism, and the dynamical picture is one of
purely longitudinal strings breaking up and producing
particles. To a high-$p_\perp$ collider physicist, the 
complete absence of jets may seem a quite radical
starting assumption, but recall that the vast majority of the
(soft-inclusive) cross section involves very small momentum
transfers. These models are typically used
e.g.\ for heavy-ion collisions and cosmic-ray air showers, for which
the small fraction of events that contain hard identifiable jets can
often be neglected (though obviously not in hard tails, such as jet-quenching studies).  
The main focus is here on the soft physics, though perturbative
contributions can be added in, e.g.\ by the introduction of a ``hard
pomeron''. In-between are tools like PHOJET~\cite{Bopp:1998rc},
DPMJET~\cite{Bopp:2005cr}, EPOS~\cite{Werner:2010aa}, and 
SIBYLL~\cite{Ahn:2009wx}, 
which contain elements of both 
languages (with EPOS adding a further component:
hydrodynamics~\cite{Werner:2007bf}). 
Note, however, that all of these models rely on string models of
hadronization and hence have some overlap with PYTHIA on that aspect of
the event modelling.

\begin{figure}[t]
\centering
\includegraphics[scale=0.32]{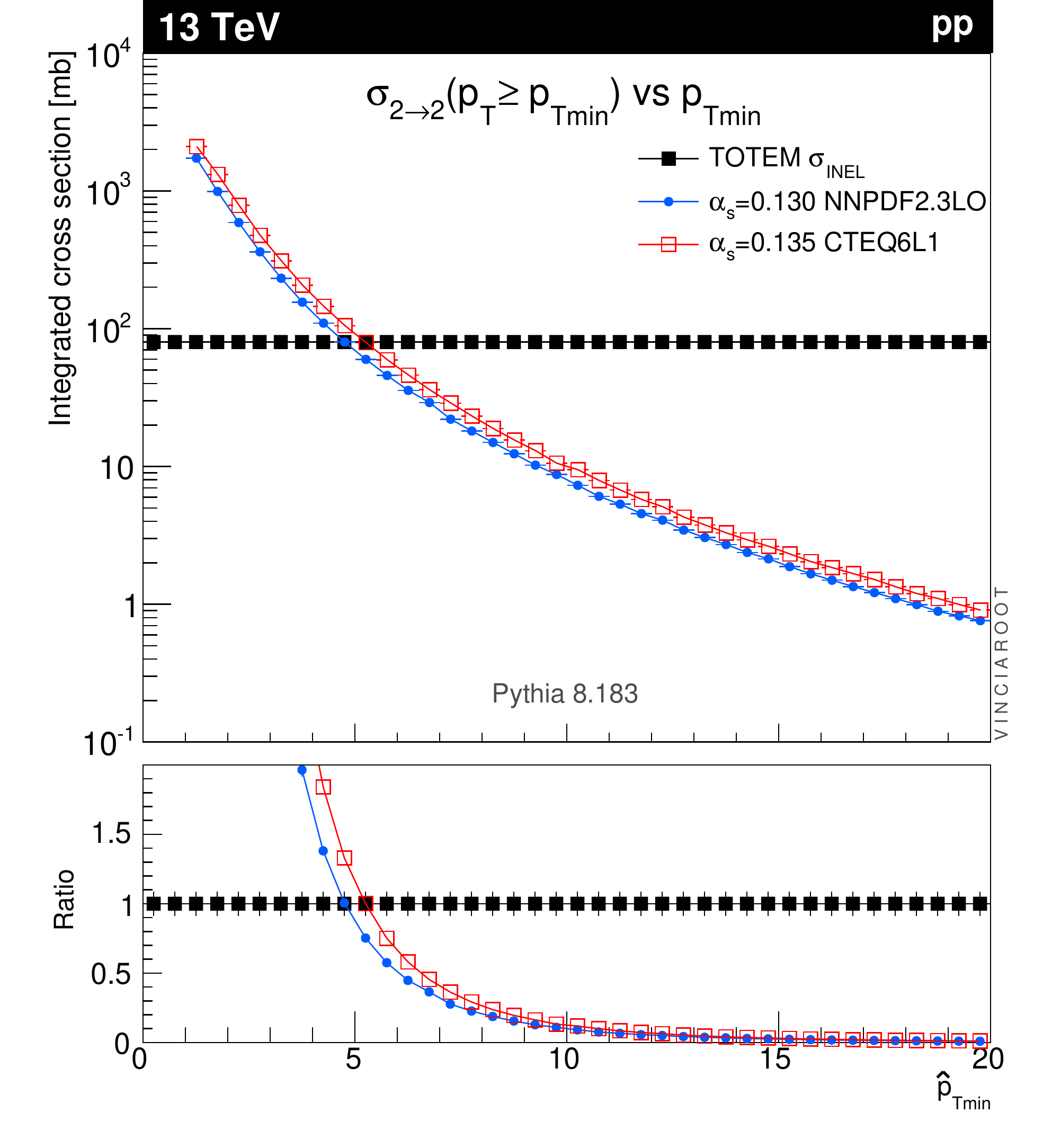}
\includegraphics[scale=0.32]{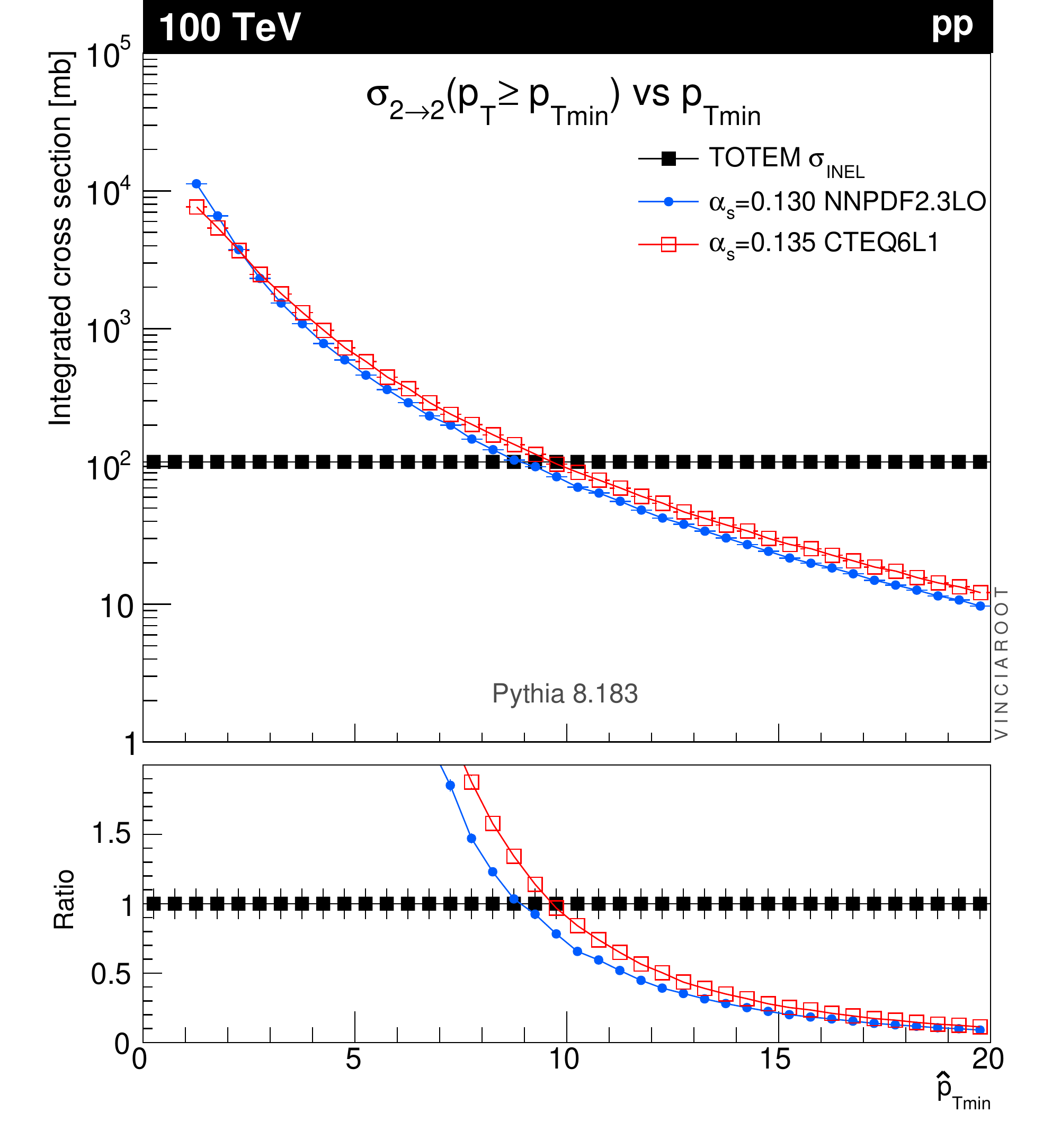}
\caption{The integrated parton-parton cross section (blue dots and
  red open squares) compared with the total proton-proton cross section
  (filled black squares) at CM energies from 13 TeV (top) to 100 TeV
  (bottom). From \cite{Skands:2014pea}.
\label{fig-exceedTotal}}
\end{figure}

Regardless of the details, any framework that attempts to combine soft
and hard QCD eventually faces the following problem:  \emph{At some point, the
perturbatively calculable (``hard'') parton-parton cross section exceeds the total 
(``hard+soft'') hadron-hadron cross section}. This is illustrated in
figure \ref{fig-exceedTotal}, for proton-proton collisions at CM
energies from 13 TeV (top pane) to 100 TeV (bottom pane). 
At each CM energy, the total (inelastic) hadron-hadron cross section (based on
\cite{Antchev:2013paa,Cudell:1996sh}) 
is shown as a horizontal line with filled black squares. The $x$ axis,
labeled $\hat{p}_{T\mathrm{min}}$, represents an arbitrary lower limit of
integration for the perturbative QCD $2\to 2$ cross sections, 
\begin{equation}
\sigma_{2\to 2}(\hat{p}_{T\mathrm{min}}) = 
  \int_{\hat{p}_{T\mathrm{min}}^2}^{s/4} d\hat{p}_{T}^2
  \frac{d\hat{\sigma}_{2\to 2}}{d\hat{p}_{T}^2}~,
\end{equation}
where the partonic $d\hat{\sigma}_{2\to 2}$ cross section is dominated
by $t$-channel gluon exchange with a
characteristic $1/\hat{t}^2$ singularity, and we have suppressed integrations over partonic $x$ fractions. The 
divergence of this partonic cross section, $1/\hat{t}^2 \sim
1/\hat{p}_{T}^4$ for low $\hat{p}_T$, augmented by 
running-coupling and low-$x$ parton-distribution effects, implies that at \emph{some}
$\hat{p}_{T\mathrm{min}}$ value, the red and blue parton-parton
cross-section curves in figure
\ref{fig-exceedTotal} must exceed the total hadron-hadron one. (The
small difference between the two curves represent different PDF and
$\alpha_s$ choices.) At the LHC at 13 TeV (top pane), the
parton-parton cross section becomes equal to the hadron-hadron one for  $\hat{p}_{T\mathrm{min}}(13~\mathrm{TeV})\sim
5~\mathrm{GeV}$, while the corresponding value at 100 TeV (bottom pane)
is $\hat{p}_{T\mathrm{min}}(100~\mathrm{TeV})\sim
10~\mathrm{GeV}$. Although these are arguably quite low scales in the
context of ``jets'', the main point is that they are still
perturbative. We do not naively expect that non-perturbative effects 
significantly reduce the $2\to 2$ cross section at $\hat{p}_T$ values
as large as 10 GeV. 

The parton- and hadron-level cross sections can be reconciled by
noting that their ratio, 
\begin{equation}
\left<n\right>(\hat{p}_{T\mathrm{min}},s) = \frac{\sigma_{2\to 2}(\hat{p}_{T\mathrm{min}},s)}{\sigma_\mathrm{INEL}(s)}~,
\end{equation}
counts how big a fraction of all (inelastic) events contain a partonic
$2\to 2$
scattering above a given $\hat{p}_{T\mathrm{min}}$, as a function of
  hadron-hadron CM energy, $\sqrt{s}$. If this fraction is
  \emph{greater} than one, it simply means that each hadron-hadron
  collision contains \emph{more} than one such partonic $2\to 2$
  scattering. Thus the idea is born: multiple perturbative
  parton-parton interactions (MPI).  

As mentioned above, MPI has historically been an essential ingredient
in the modelling of hadron-hadron collisions especially in PYTHIA (see
\cite{Sjostrand:1987su}). When augmented by impact-parameter
dependence (an aspect that goes beyond this mini-review), it
allows to describe a number of important phenomenological features,
such as the extremely wide multiplicity distributions and significant
deviations from KNO scaling~\cite{Koba:1972ng} observed already eg at
the SPS~\cite{Alner:1987wb,Ansorge:1988kn}.  Modern implementations of
partonic MPI are featured in EPOS, HERWIG++, PHOJET, PYTHIA 6 \& 8, SHERPA, and
SIBYLL 2, while QGSJET and SIBYLL 1 rely on multiple cut pomerons
(i.e., not associated with partonic jets).

\begin{figure}[t]
\centering
\begin{tabular}{rc}
\rotatebox{90}{\hspace*{1.5cm}\small$p_{\perp 0}~[\mathrm{GeV}]$} &
\includegraphics*[scale=0.63]{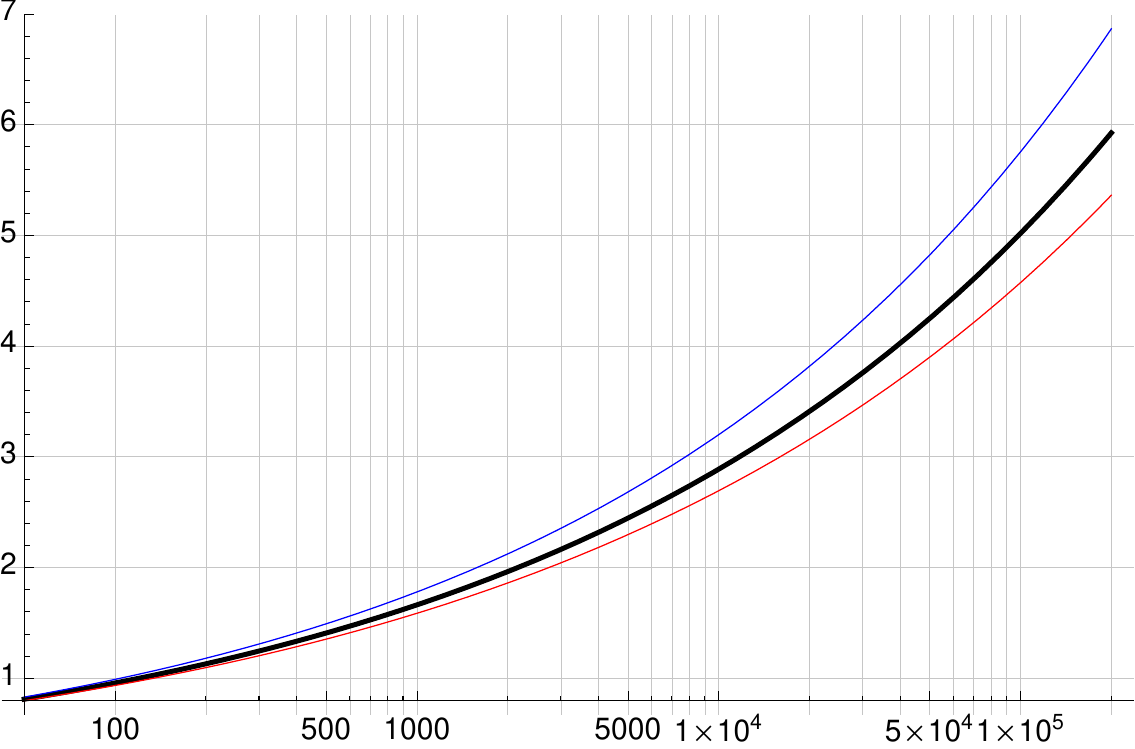}\\
&\small $E_\mathrm{CM}~[\mathrm{GeV}]$
\end{tabular}
 \caption{Scaling of
  the $p_{\perp 0}(\sqrt{s})$ soft-MPI regularisation scale in the
  Perugia 2012 tunes
  (central value and range). \label{fig-pT0}}
\end{figure}

In perturbative MPI-based models, one should be aware that 
the amount of soft MPI is sensitive to the 
PDFs at low $x$ and $Q^2$, a region which is not especially well
controlled. Physically, colour screening and/or saturation effects
should be important. In practice, one typically introduces 
an $E_\mathrm{CM}$-dependent regularisation scale, $p_{\perp
  0}(\sqrt{s})$ (of the same order and usually slightly smaller than
the $\hat{p}_{T\mathrm{min}}$ scale discussed above), which is assumed to modify 
the naive LO QCD $2\to2$ cross sections in the following way,
\begin{equation}
\frac{\mathrm{d}\sigma_{2\to 2}}{\mathrm{d}p_\perp^2} \ \propto \
 \frac{\alpha^2_s(p_\perp^2)}{p_\perp^4} \ \to \  
 \frac{\alpha^2_s(p_\perp^2+p_{\perp 0}^2)}{(p_\perp^2+p_{\perp 0}^2)^2} ~,
\end{equation}
such that the divergence for $p_\perp \to 0$ is regulated. 
For illustration, the CM energy dependence  of 
$p_{\perp 0}$  for
the so-called Perugia 2012 tunes of PYTHIA 6.4~\cite{Skands:2010ak} 
is shown in figure~\ref{fig-pT0}. 

\begin{figure}[t]
\centering
\includegraphics*[scale=0.56]{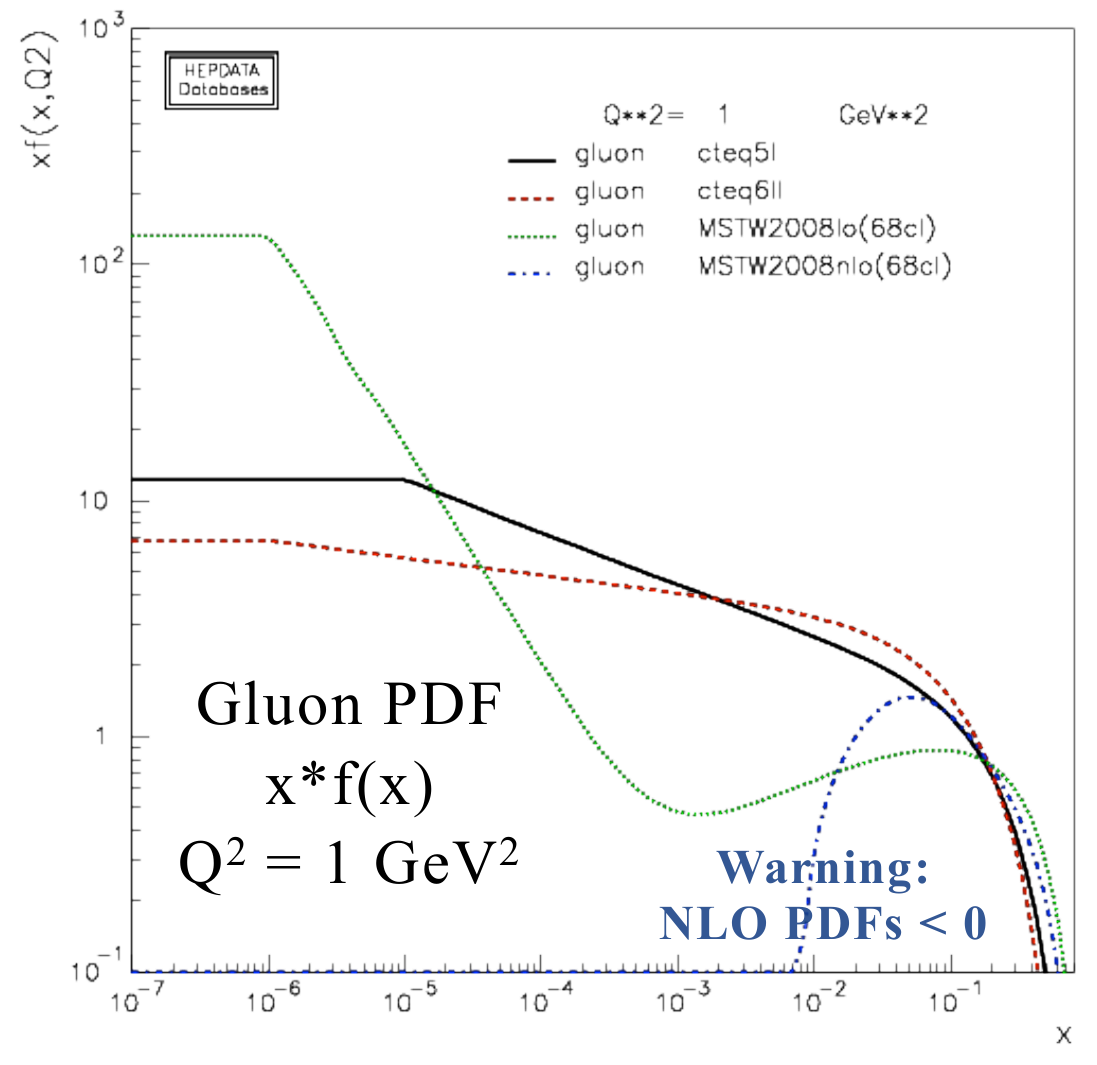} 
\caption{Behaviour of some typical PDF sets at very low
  $Q^2 = 1\,\mathrm{GeV}^2$. Plot from
  HEPDATA~\cite{Buckley:2010jn}. 
 \label{fig-PDFs}}
\end{figure}

There is then still 
a dependence on the low-$x$ behaviour of the
PDF around that scale, illustrated 
in figure~\ref{fig-PDFs} (see also \cite{Schulz:2011qy,Skands:2014pea}). 
Note the freezing of the PDFs at very low $x$ 
(only marginally relevant for $E_\mathrm{CM}\le 100\,\mathrm{TeV}$). 
Note also that NLO PDFs should not be used for MPI models, since they
are not probability densities (e.g., they can become negative, illustrated
here by the MSTW2008 NLO set~\cite{Martin:2009iq}). The Perugia 2012
tunes are based on the CTEQ6L1 LO PDF set~\cite{Pumplin:2002vw},
but include MSTW2008 LO~\cite{Martin:2009iq} and MRST
LO**~\cite{Sherstnev:2008dm} variations.  

In practice, the optimal value for $p_{\perp 0}$ (and its scaling with
the hadron-hadron CM energy) also depends on the 
IR behaviour of 
$\alpha_s$, the IR regularisation of the parton showers, and the possible
existence of other significant IR physics effects, such as colour
(re)connections, discussed below.
There is also 
an implicit dependence on the 
assumed transverse mass-density of the proton~\cite{Corke:2011yy}. 
These caveats and dependencies notwithstanding, MPI is the basic
concept driving the modelling of all inelastic non-diffractive
events, as well as the underlying event.

Turning to the specific context of the PYTHIA event generator, 
the development and support of  PYTHIA 6 has now ceased since a few years, with new
developments only being implemented in PYTHIA 8.

For reference, in PYTHIA 6, two explicit 
MPI models are available, an ``old'' one based on 
virtuality-ordered
showers~\cite{sjostrand:1985xi,Bengtsson:1986et,Bengtsson:1986hr} with
no showers off the additional MPI interactions and a comparatively
simple beam-remnant treatment~\cite{Sjostrand:1987su}, and a ``new''
one based on (interleaved) $p_\perp$-ordered showers~\cite{Sjostrand:2004ef},
including MPI showers and  
a more advanced beam-remnant treatment~\cite{Sjostrand:2004pf}. In
both cases, only partonic QCD $2\to 2$ processes are included among
the MPI (hence no multiple-$J/\psi$, multiple-$Z$, etc type
MPI processes). Most
LHC tunes (e.g., the ``Perugia'' ones~\cite{Skands:2010ak}) 
use the ``new'' $p_\perp$-ordered framework. 
Diffractive events are treated as purely non-perturbative, 
with no partonic substructure: a diffractive mass, $M$, is 
selected according to the above formulae, and the final state produced
by the diffractively excited system is modeled as a single 
hadronizing string with invariant mass $M$, stretched along the beam
axis (two strings in the case of double diffraction). 

In PYTHIA 8, the MPI model extends and
improves the $p_\perp$-ordered one from PYTHIA 6. 
The main differences are: full interleaving of final-state showers with
ISR and MPI~\cite{Corke:2010yf}; a richer mix of MPI processes,
including electroweak processes and multiple-$J/\psi$ and -$\Upsilon$
production (see the HTML manual under ``Multiparton
Interactions:processLevel''); an option to select the second MPI ``by
hand'' (see the HTML manual under ``A Second Hard Process''); 
an option for final-state 
parton-parton rescattering~\cite{Corke:2009tk} (mimicking a mild 
collective-flow effect in the context of a dilute parton system, 
see the HTML manual under
``Multiparton Interactions: Rescattering''); colour reconnections are
handled somewhat differently (see the HTML manual
and~\cite{Sjostrand:2013cya,christiansen}); and an option for an  
$x$-dependent transverse proton size~\cite{Corke:2011yy}. 

An example where the treatment in PYTHIA 8 already 
surpasses the one in PYTHIA 6 is hard diffraction (for soft diffraction,
the modelling is the same between 6 and 8, though the diffractive and 
string-fragmentation tuning parameters may of course differ). The
default modelling of hard diffraction in PYTHIA 8 is described
in~\cite{Navin:2010kk} and follows an Ingelman-Schlein
approach~\cite{Ingelman:1984ns} to introduce partonic substructure in
high-mass diffractive scattering. (``High-mass'' is defined as
corresponding to diffractive masses greater than about 10 GeV, though
this can be modified by the user, see the HTML manual under
``Diffraction''.)
This gives rise to harder $p_\perp$
spectra and diffractive jets. A novel feature of the PYTHIA 8
implementation is that hard diffractive interactions can include MPI 
(inside the Pomeron-proton system such that the rapidity gap is not
destroyed), with a rate governed by the (user-specifiable)
Pomeron-proton total cross section, $\sigma_{p\mathbb{P}}$. This
predicts that there should be an ``underlying event'' also in hard
diffractive events, which could be searched for eg in the region
``transverse'' to diffractive jets, and/or in association with
diffractive $Z$ production, which is currently being implemented in
PYTHIA 8. 

We should also note that the default parametrization of
the $pp$ and $p\bar{p}$ cross sections in PYTHIA 8 is still based on a fairly old
(1992) Donnachie-Landshof fit~\cite{Donnachie:1992ny}, with an
asymptotic behaviour $\sigma_\mathrm{TOT}\propto s^{0.08}$. This is combined with
Schuler-Sj\"ostrand parametrizations of the diffractive
components~\cite{Schuler:1993td}. More recent studies based on LHC
data~\cite{Antchev:2013iaa,Antchev:2013paa,Aad:2014dca} indicate a
steeper rise, $\propto
s^{0.096}$~\cite{Cudell:1996sh,Donnachie:2013xia}. 
In the context of PYTHIA, the
difference seems mainly to be reflected in PYTHIA predicting 
a too small elastic cross section, while the inelastic component agrees well with LHC
data. Updating the total cross sections is on the ``to-do'' list for a
future version of PYTHIA 8.

Finally, an
alternative treatment relying on the min-bias Rockefeller (MBR) model
is also available in PYTHIA 8~\cite{Ciesielski:2012mc}. 

\section{Colour Reconnections}
\label{sec-ColRec}

The issue of final-state colour reconnections (CR) is becoming
increasingly recognised as one of the main outstanding problems in
soft-inclusive hadron-hadron
physics~\cite{Rathsman:1998tp,Skands:2007zg,Sjostrand:2013cya,christiansen}, 
with significant potential implications not only for min-bias type
physics but also impacting high-$p_\perp$ precision measurements such as the
top quark mass~\cite{Skands:2007zg,Argyropoulos:2014zoa}. 

Physically, CR may reflect 
a generalisation of soft colour coherence, dense-packing, and/or collective effects
(parton-, string-, or hadron-rescattering). 
Disentangling the causes and effects of CR is likely to be a crucial
topic for soft-QCD studies to unravel during the coming years. This
will require the definition and study of CR-sensitive observables and a
detailed consideration of the interplay between PDFs, MPI, and diffractive
physics, with MPI possibly contributing to destroying rapidity gaps
in ``originally'' diffractive events, and CR possibly creating them in
``originally'' non-diffractive
ones~\cite{Edin:1995gi,Rathsman:1998tp}.

In the context of MPI models, the question of CR arises naturally when
one considers how the additional parton-parton interactions should be
represented in terms of strings or clusters fragmenting 
into hadrons. An ad-hoc solution could be to represent
the additional MPI systems as overall colour singlets, i.e., individual
strings or clusters hadronizing separately from the rest of the
event. This simple scenario existed as an option in the original PYTHIA
implementation~\cite{Sjostrand:1987su}, was the basis of the initial
HERWIG++ MPI modelling~\cite{Bahr:2008dy}, and as far as I understand is also the basis of the
modelling of the fragmentation of cut pomerons in
QGSJET~\cite{Ostapchenko:2004ss}. However, this colour structure
physically corresponds to a \emph{diffractive} colour flow (singlet
exchange), which is \emph{not} consistent with $t$-channel gluon
(colour-octet) or cut-pomeron exchange. Empirically, it also leads to
conflict with the data and e.g., produces too large forward peaks in
the charged-particle pseudorapidity spectrum (from disconnected
``diffractive-looking'' MPI systems boosted along the $z$
axis)~\cite{Gieseke:2012ft}, and predicts that the average transverse
momentum is roughly independent of charged multiplicity, in stark
contrast to observations~\cite{Sjostrand:1987su}. 

Therefore the MPI models in both PYTHIA 6 and 8 also included an
alternative (though still rather ad hoc) option for
``inserting'' the additional partons from MPI onto the string pieces
created by the primary interaction, in a way designed to minimize the
overall increase in ``string
length''~\cite{Sjostrand:1987su,Corke:2009tk,Corke:2010yf}. In studies of
Tevatron data, 
Rick Field in particular found a very strong preference for the
minimal-string-length option, resulting in the famous ``Tune A'' 
family of tunes~\cite{Field:2005sa}, which were the first to deliver a satisfactory
description of the underlying event at high energies. There remained
the physics question of \emph{how} nature arranged for this preference
to be selected. Early attempts at modelling realistic colour flow with octet
exchanges did not provide an explanation~\cite{Sjostrand:2004pf}, and
the most successful models today are still driven by string-length
miminizations~\cite{Rathsman:1998tp,Skands:2007zg,Gieseke:2012ft},
without any particularly deep understanding of the microphysics involved. An
alternative scenario is provided by EPOS, which assumes that high
string densities triggers a hydrodynamic
phase~\cite{Werner:2007bf}. Other possibilities currently under
development include the idea of ``colour ropes'~\cite{Biro:1984cf,Andersson:1991er} --- strings
carrying several units of colour charge (instead of $n$ ordinary
strings on top of each other) --- and generalised colour coherence
applied to the process of string
formation~\cite{christiansen}. Whatever the case, this is clearly a
fertile area for model building today, with potentially important
consequences, and for which we are already aware of several
sensitive observables, most importantly the evolution of $p_\perp$
spectra with charged multiplicity (and particle mass, see
e.g.~\cite{Abelev:2006cs}), but also heavy-ion inspired flow-type
observables could be revealing, the dependence of particle spectra in
the underlying event on underlying-event activity for fixed jet
$p_\perp$, and the emergence and destruction of rapidity gaps could
all carry sensitive additional information.

\section{Recent PYTHIA Tunes}\label{sec-PyTunes}
The most recent PYTHIA 8 tune is currently the Monash
2013 tune~\cite{Skands:2014pea}, which has been selected as the
new default tune since version 8.2~\cite{Sjostrand:2014zea}, replacing the earlier
Tune 4C default of Pythia 8.1~\cite{Corke:2010yf,Sjostrand:2007gs}. A summary of the main
properties are as follows: for the final-state fragmentation, it allows
10\% more strangeness in string breaks, and has somewhat softer heavy-quark (c
and b) fragmentation functions, achieving better agreement with s-,
c-, and b-sensitive observables at LEP and SLD. In the context of $pp$
collisions, it is based on a new LO NNPDF 2.3 PDF
set~\cite{Ball:2011uy,Ball:2013hta,Carrazza:2013axa}, which has a
slightly larger low-$x$ gluon than the previous default CTEQ6L1 set~\cite{cteq61},
hence the Monash 2013 tune produces more forward activity. There is
also a better agreement with the energy scaling of average min-bias
multiplicities at LHC energies, from 900 to 7000 GeV~\cite{Aamodt:2010pp}.

The tuning efforts, however, did not explicitly attempt to retune the
diffractive components, and there are still significant discrepancies
for identified-particle rates and spectra in $pp$ collisions. Those may point
towards a need for better CR models~\cite{christiansen} and/or for inclusion of other
soft/collective effects, whatever their origin. 

Important remaining open questions include dedicated tuning studies in
the context of diffraction, for instance to constrain the total
Pomeron-proton cross section, $\sigma_{p\mathbb{P}}$, which controls
the amount of MPI in hard diffractive processes, the sensitivity to
the diffractive PDFs, and dedicated tests of string-fragmentation
parameters in the specific context of diffractive final states, as
compared with LEP-tuned parameters. 

For completeness, we note that the most recent author-driven PYTHIA 6
tunes are the so-called Perugia 2012 set of
tunes~\cite{Skands:2010ak}, now superseded by the Monash 2013 tune of
PYTHIA 8. We hope to provide Perugia-like tune variations also for PYTHIA 8 tunes
in the future, though this was not done in the context of
the Monash 2013 tune.

\subsection*{Acknowledgments}
This work was supported in part by
the Research Executive Agency (REA) of the European Commission under
the Grant Agreements PITN-GA-2012-315877 (MCnet).
\bibliography{skands-isvhecri}
\end{document}